\documentclass{elsart}

\usepackage{amssymb}
\usepackage{graphicx}
\usepackage{dcolumn}
\usepackage{bm}
\usepackage{here}
\usepackage{tabularx}

\begin{document}
\begin{frontmatter}
\title{Robustness of Attractor States in Complex Networks with Scale-free Topology}
\author[Kinoshita]{Shu-ichi Kinoshita \corauthref{cor}} 
\corauth[cor]{Corresponding author.}
\ead{f01j006g@mail.cc.niigata-u.ac.jp}
\author[Iguchi]{Kazumoto Iguchi}
\author[Yamada]{Hiroaki S. Yamada}

\address[Kinoshita]{Graduate School of Science and Technology, Niigata University, 
Nishi-ku Ikarashi 2-Nochou 8050, Niigata 950-2181, Japan}
\address[Iguchi]{KazumotoIguchi Research Laboratory, 70-3 Shinhari, Hari, Anan, 
Tokushima 774-0003, Japan}
\address[Yamada]{Yamada Physics Research Laboratory, Nishi-ku Aoyama 5-7-14, Niigata 950-2002, Japan}

\begin{abstract}
We study the intrinsic properties of attractors in the Boolean dynamics in complex network
with scale-free topology, 
comparing with those of the so-called random Kauffman networks. 
We have numerically investigated the frozen and relevant nodes for each attractor, 
and the robustness of the attractors to the perturbation 
that flips the state of a single node of
attractors in the relatively small network ($N=30  \sim 200$).
It is shown that the rate of frozen nodes in the complex networks 
with scale-free topology is larger than that in
the random Kauffman model.
 Furthermore, we have found that in the complex scale-free networks
 with fluctuations of in-degree number
the attractors are more sensitive to the state flip of a highly connected node
than to the state flip of a less connected node.
\end{abstract}
\end{frontmatter}
\section{Introduction}
\label{sec01}
Dynamics of Boolean networks is often used as a model for 
genetic networks inside cells, in which the genetic states are 
represented in terms of the language of attractors
\cite{kauffman93,kauffman69}. 
A Boolean network consists of $N$ nodes,
each of which receives $k_i$ inputs such that
the degree is $k_i$.  
In the so-called {\it Kauffman model} --
a random Boolean network (RBN) model, each node  receives
a certain fixed number of inputs such that the degree is $k_i\equiv K$. 

However,  in real systems each node has a different number of inputs.
As a more realistic modeling of biological systems 
we expect that  the fluctuation of the number of input-degree is treated as 
a random variable with a probability distribution function 
such as the inverse power-law distribution 
or the exponential distribution or the Poisson distribution. 
In fact, the in-degree distribution appears to be exponential 
in $E. coil$ and to be power-law in yeast 
\cite{barabashi99,lee02,sen03,kauffman03a,skarja04,albert06,osawa02}.

In the present report, the number of in-degree $k_i$ 
at the $i-$th node is determined by the preferential attachment rule
\cite{albert00} that makes the system a complex network with scale-free
topology (SFRBN).
The details of the method for generating the networks have been given in our previous 
paper \cite{iguchi07}.
It should be stressed that in the present paper we do not deal with large networks
that the difference between the function forms of the distributions becomes clear numerically.
The network size that we deal with is at most $N=30 \sim 200$,
focusing on the intrinsic properties of Boolean dynamics in the complex networks.

\section{Model}
\label{sec01}
The initial values for the nodes are chosen randomly 
and are synchronously updated in the time steps, according to the
connectivity $\{k_i \}$ and the Boolean functions $\{f_i \}$ assigned
for each node in the network as, 
\begin{equation}
\sigma_{i}(t+1) = f_{i}(\sigma_{i_{1}}(t),\sigma_{i_{2}}(t),
 \cdots, \sigma_{i_{k_{i}}}(t)), 
\end{equation}
where $i=1,...,N$ and $\sigma_{i}  \in  \{0,1\}$ is the binary state. 
All trajectories starting at any initial state run into
a certain number of attractors(i.e. points or cycles). 
We study 
the directed RBNs, the directed SFRBNs, and
the directed SFRBNs throughout this paper.

Figure 1 shows the time-dependence of the state which constitutes an 
typical attractor in the SFRBN with $\langle k \rangle=2$.
\begin{figure}
\begin{center}
\includegraphics[scale=0.6]{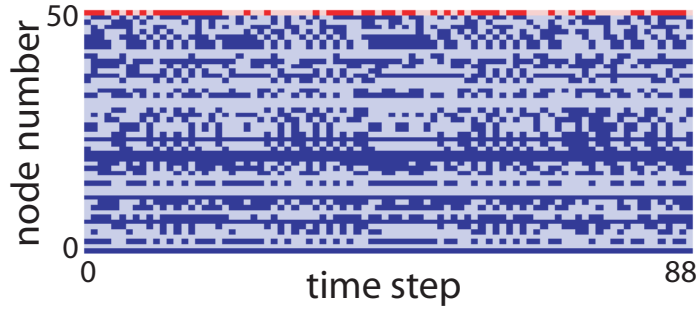}
\end{center}
\caption{
(Color online)
Space-time diagram for an typical attractor with $\ell_c=88$.
in the SFRBN with $\langle k \rangle=2$. 
The vertical axis denotes the node number that is in the high connectivity order.
The dense or gray color corresponds to the binary values of the state.
The static nodes of the attractor are frozen nodes.   
The network size is $N=50$.
}
\label{fig:1}
\end{figure}

\begin{figure}
\begin{center}
\includegraphics[scale=1.1]{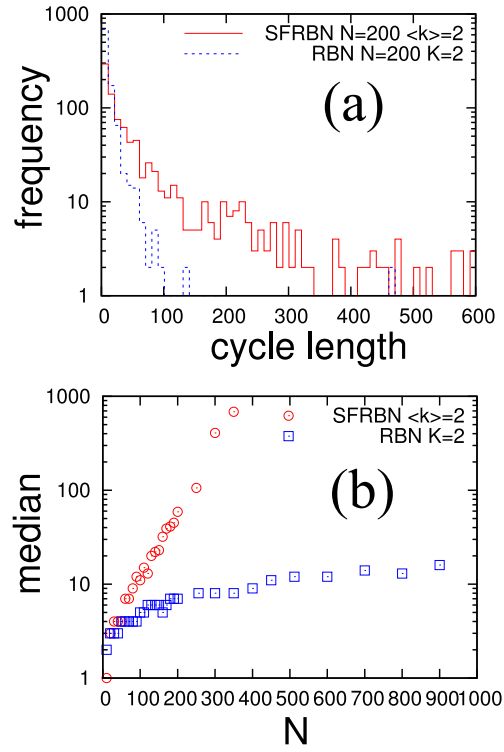}
\end{center}
\caption{
(Color online)
(a) Histogram of the length $\ell_c$ of state cycles
is shown for the RBNs and the SFRBNs, respectively, 
where the network size is $N=200$.
Each histogram is generated by $10^3$ different sets of the Boolean functions
and five different network structures.
The maximum iteration number of the Boolean dynamics is $10^5$
until the convergence to the cycle is realized.
(b) Semi-log plots of the median value $\bar{m}$ of 1000 samples of
the lengths of the state cycles
with respect to the total number $N$ of nodes for the directed RBNs and SFRBNs.
}
\label{fig:1}
\end{figure}
Fig. 2(a) shows the histogram of the length $\ell_c$ of the attractors
in  the  RBN with $K = 2$ and that in the SFRBN  where the average degree 
of nodes $\langle k \rangle = 2$.
Fig. 2(b) shows the median value $\bar{m}$ of the distribution of state cycle
lengths with respect to $N$, the total number of nodes. 
Apparently, the distribution of the attractor lengths in the SFRBN is much wider
than that in the RBN, 
and the attractor length has longer period than that in the RBN.  
This is directly related to the diversity of attractors in the SFRBNs,
which is of great importance for the stability of living cells.
We investigated the function form $\bar{m}(N)$ in more detail 
in the previous paper \cite{iguchi07}, 
and found that the function form $\bar{m}(N)$
asymptotically changes from the algebraic type  $\bar{m}(N) \propto
N^\alpha$ to the exponential one 
as the average degree $\langle k \rangle $ goes to $\langle k \rangle =2$.

Here we have a question: {\it How does the characteristics of the attractors depend on the 
network topology?} 
In the present paper we study the difference between attractors 
in the RBN and the SFRBN without a bias,
focusing on the frozen nodes and the robustness of attractors 
against the external perturbations.

\section{Frozen Nodes of Attractors}
In this section, we investigate the so-called frozen nodes of attractors 
 whose values remain constant through a given trajectory of the attractors 
\cite{kauffman93}. 
Frozen nodes arise through canalizing the Boolean functions and the homogeneity bias.

We count  the number of frozen nodes $N_f$ for each attractor and plot
the histograms for some cases in Fig.3. 
The remarkably different peak structure exists between 
the cases in the SFRBN of $\langle k \rangle=2$ and the RBN of $K=2$. 
The distributions in the SFRBN have a peak around $N_f \sim N/2$,
while the distributions in the RBN are broad with a peak at $N_f=N$. 
Note that $N_f=N$ corresponds to the point attractors that all nodes are 
frozen.

\begin{figure}
\begin{center}
\includegraphics[scale=1.1]{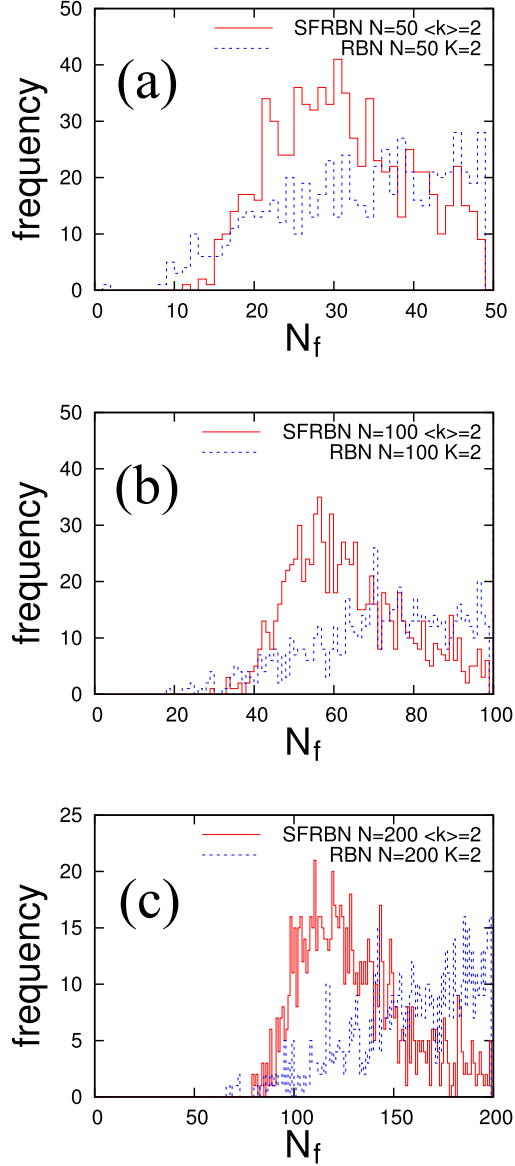}
\end{center}
\caption{
(a) Histograms of the number of frozen nodes $N_f$ for 1000 attractors 
in RBNs with $K=2$ and SFRBNs with $\langle k \rangle=2$.  The network size is (a)$N=50$,
(b)$N=100$ and  (c)$N=200$.
The scale out data at $N_f=N$ are not shown in the figures.
}
\label{fig:1}
\end{figure}

The fact that $N_f$ is relatively smaller
 in the SFRBN
would make the attractor period larger than that in the RBN, as seen in Fig.2.
However, role of the frozen nodes is not so clear in the network because all frozen nodes
also are connected to the network and might influence on the attractors.
In the next section, we investigate the significance of each node 
in the respective attractor.

\section{Robustness of Attractors to State Flip}
One of the important properties in the scale-free topology is the
existence of the highly connected hub node 
as seen in the yeast synthetic network and so on.
 In this section, we investigate the robustness of attractors to an external perturbation 
caused by an inversion of the binary state of a single node.
We consider an attractor of period $\ell_c$ and flip
the state  of the single node at time $t \in [1,\ell_c ]$ as a
perturbation. 
The perturbation to the trajectory of the attractor may
leap from  the trajectory of the original attractor to another one, i.e.
the attractor shift. 
The high homeostatic stability implies low reachability among different attractors.

We investigate the probability $R_s$
that the attractor remains in the original attractor under the inversion
of the single node state \cite{kauffman69,aldana03}, 
which is
the rate returning to the original attractor under the inversion.
Here we call the rate the {\it robustness} of the attractor.

\begin{figure}[h]
\begin{center}
\includegraphics[width=0.7\linewidth]{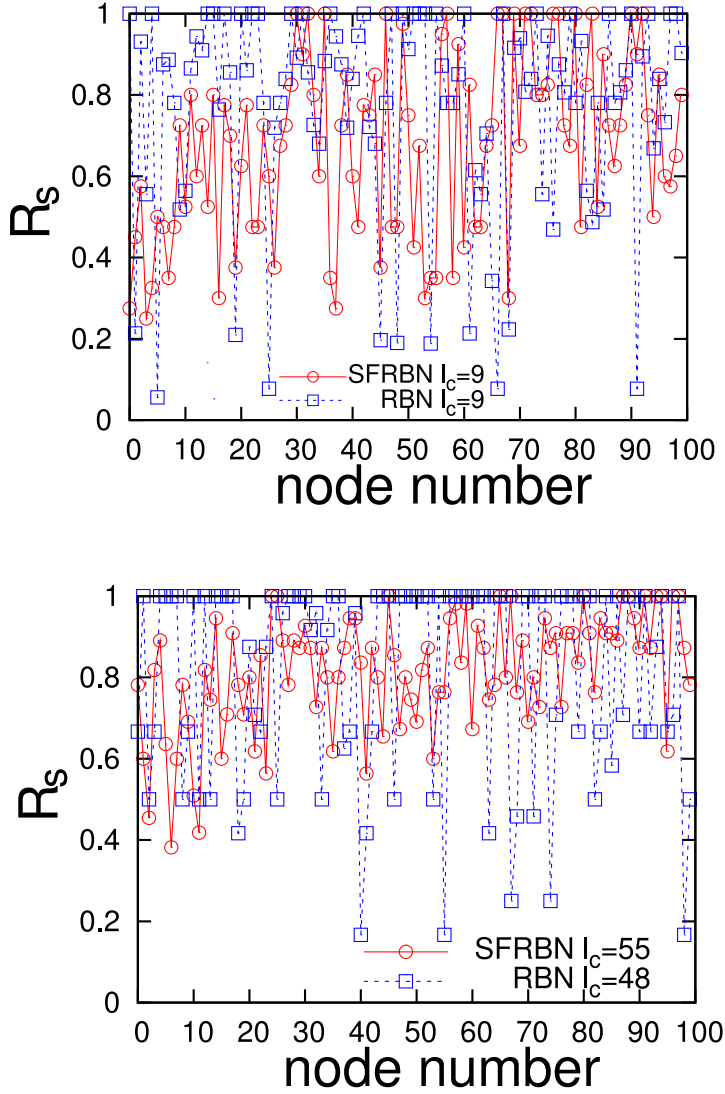}
\end{center}
\caption{
The rate $R_s$ of returning to the original attractor as a function of numbered nodes 
in the order of the number of in-degree $k_i$. 
We used $N=100$ in both the SFRBN  with $\langle k \rangle=2$ and the RBN with $K=2$.
The periods of the attractors are 
(a)$\ell_c=9$ in the SFRBN and $\ell_c=9$ in the RBN, 
(b)$\ell_c=55$ in the SFRBN and $\ell_c=48$ in the RBN, 
respectively.  
The horizontal axis "Node number" denotes the node number in the order of the
 number of input-degree.}
\label{myid:fig:sample}
\end{figure}

Figure 4 shows the robustness $R_s$ of attractors with $\ell_c=55$ in
the SFRBN and $\ell_c=48$ in the RBN, respectively.
It follows that in the SFRBN the number of "active nodes" ($R_s <1$) is much
more larger than that in  the RBN. 
On the other hand, in the RBN the perturbation to the
active nodes influences  effectively the shift of attractor ($R_s <
0.6$)  although the number of active nodes is not so many
\cite{kinoshita07}.
As a result, in the SFRBN the perturbation to the highly connected hubs 
may give rise to the attractor shift, comparing with the one to the less connected nodes.

\begin{figure}[h]
\begin{center}
\includegraphics[width=0.5\linewidth]{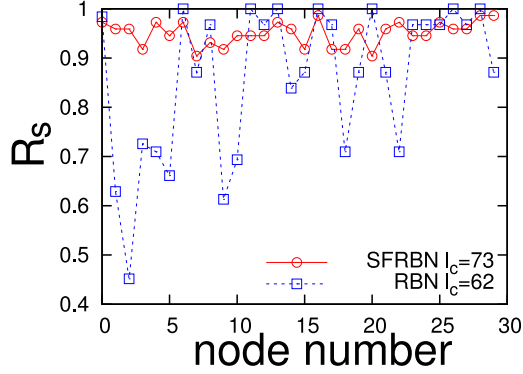}
\end{center}
\caption{
Robustness $R_s$ for some attractors of 
the SFRBN  $\langle k \rangle = 4$ and the RBN with $K=4$
in the network size of $N=30$.
The periods of attractors  are 
$\ell_c=73$ in the SFRBN and $\ell_c=62$ in the RBN.
}
\label{myid:fig:sample}
\end{figure}

The robustness of  some attractors in the SFRBN with 
$\langle k \rangle=4$ and in the RBN with $K=4$ is given in Fig.5. 
Although the whole structure is almost similar to that for the cases in Fig. 4,
it is found that the 
effect of inversion of the single site state on the attractor shift is relatively small 
compared to the cases of $\langle k \rangle=2$ and $K=2$.
There is a tendency that
the attractors become more robust to the perturbation
as the average number of input-degree $\langle k \rangle$ increases.

\section{Summary and Discussion}
\label{sec06}
In summary, we have studied the Boolean dynamics of the Kauffman model with
the directed SFRBN, comparing with the ones with the directed RBN
for the relatively small network size.
In this study we investigated some intrinsic properties of attractors 
between the RBNs and the SFRBNs,
focusing on the frozen nodes and the robustness to a perturbation. 
The obtained results are as follows. 

(i) The number of frozen nodes in the SFRBN is smaller than that in the RBN 
and 
the property reflects on the much more widely distributed attractor lengths. 

(ii) The perturbation to the highly connected hubs may give rise to the attractor shift 
in comparison to the less connected nodes.

(iii) The attractors becomes more robust to the perturbation
as the average number of input degree $\langle k \rangle$ increases.

Although in this report we did not show the details of the attractor shifts by the perturbation, 
we will present the details of the numerical results for the diagram of
transition among the attractors and the robustness to perturbation
in our forthcoming paper \cite{kinoshita07}.

Robustness against genetic mutations and environmental perturbations is 
one of the universal features of biological systems.
And the robustness is important for understanding
 evolutionary processes and homeostasis of gene regulatory networks
\cite{yam04,monte05,kauffman03}.  
However, 
in this report we investigated only robustness to the single site inversion.
For the purpose of the study, the other robustness of the SFRBN might become significant; 
for instance,  robustness of attractors to the change of Boolean functions 
and to the breakdown of the network structures.
We expect that such a study on the robustness of attractors 
provide some insights into important biological phenomena such as
cellular homeostasis and apotosis.


Actually Aldana {\it et al}  investigated the small SFRBN ($N\sim 15-20$) and found that 
the robustness of the ordered phase to the network damage 
is lower than that in the RBN \cite{aldana03}
The result implies that there exists a possibility of evolution though 
mutation even in the ordered phase, despite of the Kauffman's conjecture 
that life evolves in "edge of chaos".

Moreover, recently, an interesting network model, the so called {\it feedback network} 
has been proposed by White {\it et al} \cite{white06}. 
The feedback networks can be a good model for describing
the autocatalytic chemical reactions and the kinship, and so on, 
because the node selection, the search distance and the search path of 
networks are controlled by the attachment, 
the distance decay and the cycle formation parameters. 
It is interesting to investigate the features of 
the Boolean dynamics in  the feedback networks
from the point of view of frozen nodes and robustness \cite{kinoshita07}.

\section*{Acknowledgments}
We would like to thank Dr. Jun Hidaka for collecting many relevant papers
on the Kauffman model and the related topics.
S.K. would like to thank  professor
M. Goda for encouragement on this study.
K. I. would like to thank Kazuko Iguchi for her continuous
financial support and encouragement.


\end{document}